\documentclass[%
 reprint,
 amsmath,amssymb,
 aps,
]{revtex4-2}

\usepackage{graphicx}
\usepackage{dcolumn}
\usepackage{bm}
\usepackage{braket}

\begin{document}

\preprint{APS/123-QED}

\title{Introduction to Majorana Zero Modes in a Kitaev Chain} 

\author{Sihao Huang}
\affiliation{Department of Physics, Massachusetts Institute of Technology, Cambridge, MA 02139, USA}
\date{\today}

\begin{abstract}

Majorana zero modes (MZMs) realize a representation of non-abelian braid groups that enable topological quantum computation, wherein the storage and manipulation of information occur in decoherence-free degrees of freedom. This paper is meant to build off existing texts in the topic \cite{top_cond, herviou_topological_2017} to serve as an introduction to MZMs at the undergraduate level, providing an overview of many important concepts in condensed matter physics, such as the Bogoliubov-de Gennes formalism and topological phases. We provide an overview of the motivation behind finding MZMs and discuss how the nature of these modes can provide an elegant solution to various challenges in quantum computing. We then show how MZMs can appear in a toy model known as a Kitaev wire by solving its spectrum, and conclude with a discussion on the conditions under which they arise in a one-dimensional chain. 

\end{abstract}

\maketitle

\section{Introduction}
\label{sec:introduction}

Over the past two decades, many physical candidates for building quantum computers, such as superconducting qubits \cite{kjaergaard_superconducting_2020}, trapped ions \cite{bruzewicz_trapped-ion_2019}, and Rydberg atoms \cite{bernien_probing_2017} have emerged. While each of these modalities is able to implement a universal quantum gate set and have demonstrated simple algorithms, they are prone to two issues. First, qubits decohere over time through both energy decay --- corresponding to a classical bit-flip error from $\ket{1}$ to $\ket{0}$ --- and dephasing, wherein the relative phases between different states of the wavefunction are smeared out. Second, operations that enable information processing, which in general are unitary transformations that act on the qubits, $\ket{\psi} \to U(\theta_1, \theta_2, \ldots)\ket{\psi}$, have less than ideal fidelity (e.g. over- and under-rotations of $\theta_1 \pm \delta$) and can accumulate errors over repeated application.

While fault-tolerant computation is possible with the use of error-correcting codes, thousands of physical qubits are needed to implement a single fault-tolerant logical qubit using current technologies. As a result, schemes have been proposed to build fault tolerance at the physical level, with Majorana qubits being a prominent example.

To address the problem of decoherence, we need to prevent both bit-flip errors and dephasing. The former can be mitigated by encoding the $\ket{0}$ and $\ket{1}$ states with unoccupied and occupied electron sites respectively. Due to charge conservation, the only way for an error to occur is at two locations simultaneously, which can be suppressed by placing the two sites physically apart from each other. The latter challenge --- dephasing --- occurs due to fluctuations $\delta n$ around the occupation number $N$ of the qubit, which yields random fluctuations in its transition frequency \cite{schuster_ac_2005}. This corresponds to the term $a^\dagger a$, where $a$ and $a^\dagger$ are the fermionic creation and annihilation operators. These fluctuations arise due to couplings to electromagnetic modes in the environment, such as those driven by material defects or by the measurement field. The key insight Kitaev provided is that this can be solved by ``splitting" the fermionic site into two Majorana modes, which can then be spatially separated. 

A Majorana fermion is a particle that is its own antiparticle. In the language of second quantization, this means that $\gamma = \gamma^\dagger$, i.e. the fermionic operator $\gamma$ squares to 1. The creation and annihilation operators can be written as a superposition of two Majorana operators,

\begin{equation}
\label{eqn:superposition}
    a^\dagger = \frac{1}{\sqrt{2}}(\gamma_1 + i \gamma_2), \:
    a = \frac{1}{\sqrt{2}}(\gamma_1 - i \gamma_2).
\end{equation}

As such, they also fulfill the commutation identity $\{\gamma_i, \gamma_j\} = 2 \delta_{ij}$. The task at hand is to physically separate the two Majorana modes, $\gamma_{2j}$ and $\gamma_{2j-1}$, that make up a single fermionic mode, such that phase errors corresponding to $a^\dagger_j a_j = (1 + i \gamma_{2j-1}\gamma_{2j})/2$ are unlikely to occur. Put together, these properties would make the Majorana qubit immune to decoherence.

These Majoranas fermions can arise as quasiparticles in superconducting systems, which we will explore in a one-dimensional chain first proposed by Kitaev \cite{kitaev_unpaired_2001} in Sec.~\ref{sec:kitaev_chain}. We will see that they are bound to zero energy, making them Majorana zero modes --- a more apt name given that they no longer obey fermionic statistics --- where $[H, \gamma_i] = 0$, with $H$ being the Hamiltonian of the system (more realistically, this condition is relaxed to  $[H, \gamma_i] \approx e^{-x/\xi}$ \cite{sarma_majorana_2015}, where $x$ is the distance between the MZMs and $\xi$ is the correlation length of the Hamiltonian, as discussed in Sec.~\ref{subsec:stability}). Finally, we will see why they obey non-abelian statistics that enable the implementation of braid operations in Sec.~\ref{sec:non-abelian}. This solves the final piece of the puzzle, where qubit operations are now intrinsically fault-tolerant due to their topological properties.

\section{The Kitaev Chain}
\label{sec:kitaev_chain}

\begin{figure}
\includegraphics[width=0.5\textwidth]{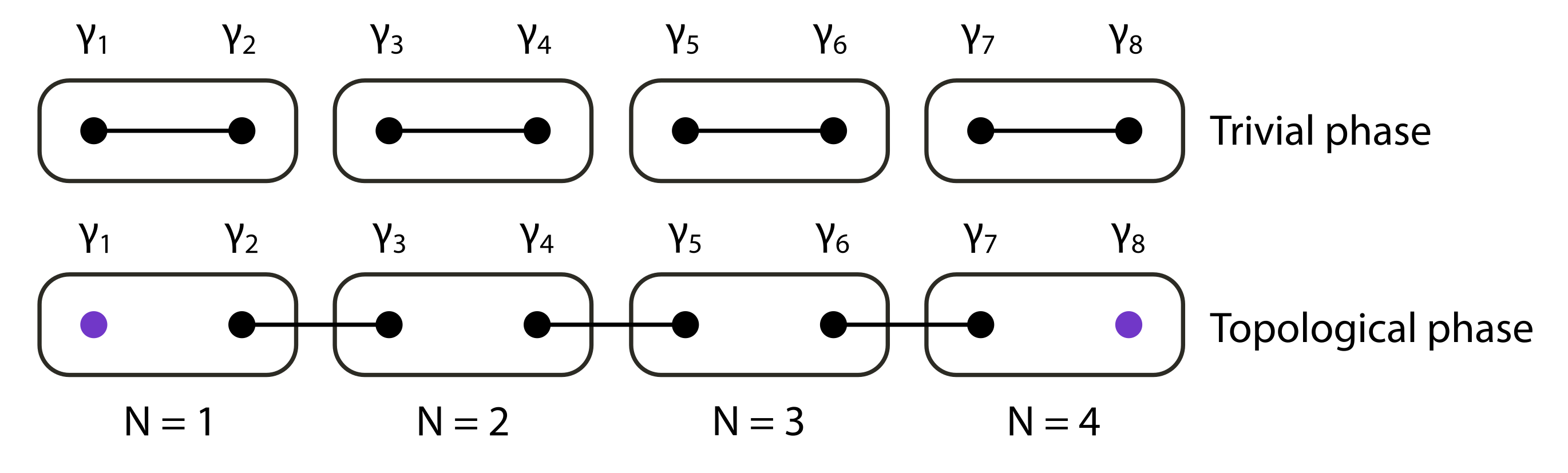}
\caption{\label{fig:chain} Two ways to pair the the Majorana modes in a Kitaev chain. Each fermionic site (labeled $N = 1 \ldots 4$) is comprised of two Majorana modes, $\gamma_{2N}$ and $\gamma_{2N+1}$. In the trivial phase, all the Majorana pairs are paired, while in the topological phase, the pairing is offset by one, leaving unpaired modes $\gamma_1$ and $\gamma_8$ at the two end.}
\end{figure}


To produce unpaired MZMs, Kitaev considered a basic 1D toy model with N fermionic sites, each corresponding to creation operators $a_n^\dagger$ \cite{kitaev_unpaired_2001}. These can each be written as a linear superposition of Majorana operators using Eq.~\ref{eqn:superposition}. Since Majoranas do not occur on their own in condensed matter systems, fermionic parity must be respected in the overall Hamiltonian --- that is, we must have an even number of Majoranas. As such, the Majorana operators must come in pairs, yielding a general quadratic Hamiltonian of the form

\begin{equation}
    H = \frac{i}{4} \sum_{l,m} A_{lm}\gamma_l \gamma_m,
\end{equation}

where Hermiticity enforces the condition that $A^\dagger = A$. A second and more subtle condition is that the $U(1)$ gauge symmetry of the electron, $a_j \to e^{i\phi} a_j$, must be broken to a $\operatorname{Z}_2$ symmetry $a_j \to -a_j$ such that transformations do not mix the Majorana operators. This motivates the use of superconducting systems, where the pairing of the electrons due to phonon-mediate interactions effectively transmutes them into two species of bosons --- particles and holes --- that give rise to $U(1) \mapsto \operatorname{Z}_2$ symmetry breaking. These conditions are fulfilled by the Hamiltonian for the Kitaev chain, given by

\begin{equation}
\label{eqn:kitaev}
\begin{aligned}
    H = \sum^{N}_{n=1} [ \underbrace{-\mu(a^\dagger_n a_n -  \frac{1}{2})}_\text{Occupation energy} \underbrace{-w(a^\dagger_{n}a_{n+1} + a^\dagger_{n+1}a_n)}_\text{Particle hopping} \\ + \underbrace{\Delta a_n a_{n+1} + \Delta^*a^{\dagger}_{n+1} a^{\dagger}_n}_\text{Particle pairing}].
    \end{aligned}
\end{equation}

The first parameter $\mu$ simply corresponds to the occupation energy of each site, which can host either $0$ or $1$ electron. The second parameter $w$ is the hopping energy, with terms $a^\dagger_n a_{n+1}$ and $a^\dagger_{n+1}a_n$ corresponding to particle motion towards the left and right ends of the chain. Finally, the third parameter, $\Delta = |\Delta| e^{i\theta}$, corresponds to the induced superconducting gap. This is the energy it takes to break up the aforementioned Cooper pairs, created by pairwise operators $a^{\dagger}_n a^{\dagger}_{n+1}$ and annihilated by $a_n a_{n+1}$, with $\theta$ being the spontaneously chosen coherent phase of the superconductor \cite{ostnell_quantum_2019}. 

It is also important to note that only one spin direction exists in this Hamiltonian --- the reason for which will be clear towards the end of Sec.~\ref{subsec:conditions}. Physically, the superconducting terms in the Hamiltonian can be achieved by depositing a nanowire onto a p-wave superconductor, which enables Cooper pairs to tunnel into the chain via the proximity effect. The p-wave nature refers to the $\ell = 1$ total angular momentum of the electron pairs, which allows for a uniform spin direction (e.g. $\ket{\uparrow\uparrow}$).

This Hamiltonian has two interesting limits. First, if we set the hopping and superconducting terms $|\Delta| = w = 0$ and let $\mu < 0$, we obtain a system of $N$ isolated fermionic sites, each with an occupation energy $\mu$,

\begin{equation}
    H = -\mu \sum^{N}_{n = 1} a_n^{\dagger} a_n.
\end{equation}

To see what this looks like in the Majorana basis, we can rewrite this with Majorana operators $\gamma$ using Eq.~\ref{eqn:superposition}:

\begin{equation}
\begin{aligned}
    H = -\mu \sum^{N}_{n = 1} \frac{1}{2} (\gamma_{2n-1} - i \gamma_{2n})(\gamma_{2n-1} + i \gamma_{2n}) \\ 
    = -\frac{\mu}{2} \sum^{N}_{n = 1} i \gamma_{2n-1} \gamma_{2n} - i\gamma_{2n} \gamma_{2n-1} + 2 \\
    = - \mu \sum^{N}_{n = 1} i \gamma_{2n-1} \gamma_{2n} + 1,
\end{aligned}
\end{equation}

where we used $\gamma^2 = 1$ in the second line and the anticommutator $\{\gamma_i, \gamma_j\} = 2 \delta_{ij}$ in the third. This simply corresponds to an even pairing of Majoranas into fermions, as indicated by the trivial phase in Fig.~\ref{fig:chain}.

Something different happens if when we consider the case where $|\Delta| = w > 0$, and the on-site energy $\mu$ vanishes. In this case, it is convenient to absorb the superconducting phase $\theta$ into the Majorana operators:

\begin{equation}
\begin{aligned}
\label{eqn:majorana_operators}
    a_n = \frac{1}{2}(e^{- i\theta / 2} \gamma_{2n} + e^{i\theta / 2} \gamma_{2n-1}), \\
    a_n^{\dagger} = \frac{1}{2}(e^{i\theta / 2} \gamma_{2n} - e^{-i\theta / 2} \gamma_{2n-1}).
\end{aligned}
\end{equation}

We can thus take $\Delta$ to be real and positive for the rest of the paper. Plugging this into Eq.~\ref{eqn:kitaev} and again using operator identities for $\gamma$, we obtain

\begin{flalign}
\begin{aligned}
\label{eqn:phase2}
    & H = - w \sum^{N-1}_{n = 1} (a_n^\dagger a_{n+1} + a^\dagger_{n+1} a_n) + (e^{i\theta} a_n a_{n+1} + e^{-i\theta} a_{n+1} a_n) \\
    & = -\frac{w}{4} \sum^{N-1}_{n - 1} [(\gamma_{2n} - i \gamma_{2n-1}) (\gamma_{2(n+1)} - i \gamma_{2(n+1)-1}) \\ & + (\gamma_{2(n+1)} - i \gamma_{2(n+1)-1}) (\gamma_{2n} - i \gamma_{2n-1})] \\ & + [(\gamma_{2n} + i \gamma_{2n-1})(\gamma_{2(n+1)} + i \gamma_{2(n+1)-1})  \\ & + (\gamma_{2n} - i \gamma_{2n-1})(\gamma_{2(n+1)} - i \gamma_{2(n+1)-1})] \\ & = - iw \sum^{N - 1}_{n = 1} \gamma_{2n} \gamma_{2(n+1) - 1},
\end{aligned}
\end{flalign}

where $\gamma_{2n}, \gamma_{2n+1}$ from different sites are now paired together. This new Hamiltonian can be diagnoalized by introducing $N - 1$ fermionic operators $\tilde{a}_n = (\gamma_{2n} - i\gamma_{2n+1})/2$, such that

\begin{flalign}
\begin{aligned}
    & H = -iw \sum^{N - 1}_{n = 1} \gamma_{2n} \gamma_{2(n+1) - 1} 
    = -iw \sum^{N - 1}_{n = 1} i(\tilde{a}_n + \tilde{a}_n^\dagger) (\tilde{a}_n - \tilde{a}_n^\dagger) \\ 
    & = w \sum^{N - 1}_{n = 1} \tilde{a}_n \tilde{a}_n + \tilde{a}^\dagger_n \tilde{a}^\dagger_n + \tilde{a}_n \tilde{a}_n^\dagger + \tilde{a}^\dagger_n \tilde{a}_n 
    = 2w \sum^{N - 1}_{n = 1} (\tilde{a}_n^\dagger \tilde{a}_n - \frac{1}{2}),
\end{aligned}
\end{flalign}

where we used the anticommutator for fermions $\{\tilde{a}_i, \tilde{a}_j\} = \delta_{ij}$ and the fact that $\tilde{a}_n \tilde{a}_n$ and $ \tilde{a}^\dagger_n \tilde{a}^\dagger_n$ both leave the system invariant, as each site only has an occupancy of $0$ or $1$. 

In this case, we started with $N$ fermionic operators in the Hamiltonian but are now left with $N-1$. This is because the Majorana modes are paired with odd parity (as indicated in Fig.~\ref{fig:chain}), so the two modes at the ends do not appear in the Hamiltonian. As such, $[H, \gamma_1] = [H, \gamma_{2N}] = 0$, and they have no energy. This gives rise to the term \textit{Majorana zero modes}. 

Finally, we can define a new fermionic operator $\tilde{a}_0 = (\gamma_1 + i \gamma_{2N})/2$, composed of MZMs at the two ends of the chain. This enables us to define the degenerate computational basis states $\tilde{a}_0 \ket{0} = 0$ and $\tilde{a}_0^\dagger \ket{0} = \ket{1}$, in which we can store and manipulate quantum information. As desired, we have successfully split the qubit into two spatially separated modes.

\subsection{Bulk spectrum of the Kitaev chain}
\label{subsec:stability}

We have thus far shown that the Kitaev chain can host unpaired MZMs in one of its phases. While it may seem to exist only under very specific conditions --- given that we set $\mu = 0$, and disorder is nearly impossible to avoid in a physical system --- we will show that the modes are actually stable over a wide range of parameters. Specifically, they persist as long as there is a finite gap in the bulk spectrum.

To compute the spectrum, we will employ the Bogoliubov-de Gennes formalism, which enables us to solve our quadratic Hamiltonian. Intuitively, we will see that Eq.~\ref{eqn:kitaev} can be reduced to a $2 \times 2$ matrix $H_{BdG}$ in the space of particles and holes (corresponding to fermionic creation and annihilation operators). This can be diagonalized by rotating it with a matrix $R$, which also acts upon the vector of particles and holes to mix them. This superposition of particles and holes yields a new operator $b = u a + v a^\dagger$ where $u, v \in \mathbb{C}$ (and the corresponding $b^\dagger$) --- so-called ``quasiparticles" which, as we will see, correspond to the two branches of the spectrum.

More precisely, we first break down the original Hamiltonian into $H = \Psi^\dagger H_{BdG} \Psi$, where $\Psi$ is a vector of creation and annihilation operators in the form of $(a_1, \ldots a_N, a_1^\dagger, \ldots, a_N^\dagger)^T$ of dimension $2N$, sometimes also known as a Nambu spinor \cite{herviou_topological_2017}. This permits us to write the $2N \times 2N$ matrix $H_{BdG}$ in particle and hole space using $2 \times 2$ Pauli matrices and basis states $\ket{n} = (0, 0, \ldots, n, \ldots, 0)$ corresponding to the $n$-th site of the chain. For example, the chemical potential term in the original Hamiltonian can be written as

\begin{flalign}
\begin{aligned}
    & - \mu (a_n^\dagger a_n - \frac{1}{2}) = -\frac{\mu}{2} (2 a^\dagger_n a_n - 1) = -\frac{\mu}{2} (-a_n a_n^\dagger + a_n^\dagger a_n) \\ & = - \frac{\mu}{2} \Psi^\dagger \begin{pmatrix} - \ket{n}\bra{n} & 0 \\ 0 & \ket{n}\bra{n}
\end{pmatrix} \Psi.
\end{aligned}
\end{flalign}

We can transform the full Hamiltonian this way, obtaining 

\begin{equation}
\label{eqn:bdg}
\begin{aligned}
    H_{BdG} =  -\sum_n  (w\sigma_z+i\Delta\sigma_y)\ket{n}\bra{n-1} + \\ (w\sigma_z-i\Delta^*\sigma_y)\ket{n-1}\bra{n} + \mu \sigma_z \ket{n}\bra{n}.
\end{aligned}
\end{equation}

\begin{figure}
\includegraphics[width=0.5\textwidth]{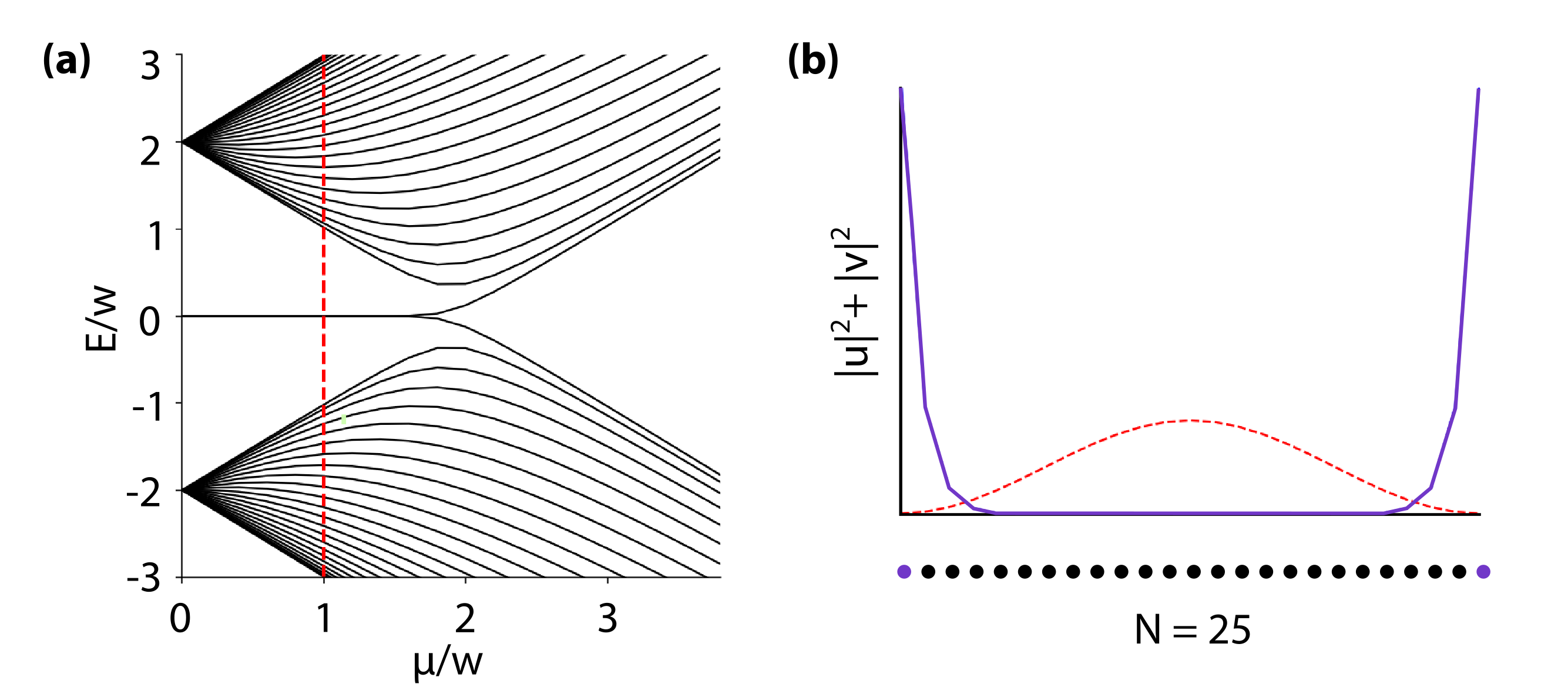}
\caption{\label{fig:spectrum} Panel \textbf{(a)} shows the spectrum of a Kitaev chain with $N = 25$ as visualized in Sau et al. \cite{sau_kitaev_nodate}, which is symmetric about $E = 0$ as predicted. Starting with $\Delta = w$ and setting $\mu = 0$, we see a zero energy mode surrounded by the bulk on both sides, which remains stable until $\mu/w \approx 2$ when the bulk gap closes. Indeed, in the wavefunction picture shown in \textbf{(b)} (plotted for $\mu/w = 1$), the MZMs are well-localized on the two ends of the chain. As $\mu$ is increased, they are slowly delocalized until they merge into a fermionic mode and split in the spectrum.}
\end{figure}

This new $H_{BdG}$ acts on basis states $\ket{n} \otimes \ket{\sigma}$ where $\sigma = \pm 1$ corresponds to the electron and hole states. The Pauli matrices act on $\ket{\sigma}$ states, with the $\sigma_y$ terms mixing the electrons and holes. Note immediately we have particle-hole symmetry in the system as the annihilation operators for the particles can be viewed as creation operators for holes, and vice-versa. Defining the particle-hole symmetry operator $P = \sigma_x K$ (where $K$ indicates complex conjugation), we see that $P H_{BdG} P^{-1} = -H_{BdG}$. As such, the spectrum has to be symmetric around zero energy, as shown in Fig.~\ref{fig:spectrum} (a).

The next step is to take a Fourier transform of the spectrum. Since we are interested in the bulk for now, let us close the chain with periodic boundary conditions such that $H_{BdG}$ has translational symmetry $\ket{n} \to \ket{n+1}$ \cite{erkensten_majorana_2019}. We can rewrite the states in momentum space using Bloch's theorem, which states that solutions to Schrodinger's equation in a periodic potential can be broken down into the product of a plane wave and a periodic function:

\begin{equation}
    \ket{k} = \frac{1}{\sqrt{N}} \sum^{N}_{n = 1} e^{-ikn}\ket{n},
\end{equation}

which moves us into the momentum space $k$. Periodic boundary conditions enforce $k = 2\pi p/N$ where $p$ ranges from $0$ to $N-1$, and $k \in (-\pi, \pi]$, allowing us to rewrite the BdG Hamiltonian 

\begin{flalign}
\begin{aligned}
    & H_{BdG} = -\sum_n \sum_k \sum_{k'} [(w\sigma_z + i\Delta \sigma_y)e^{i(k-k')n}e^{-ik} \\ & + (w\sigma_z - i\Delta \sigma_y)e^{-i(k-k')n}e^{ik} ] \ket{k}\bra{k'} + \\ & \frac{\mu \sigma_z}{N} e^{-i(k-k')n}\ket{k}\bra{k'},
\end{aligned}
\end{flalign}

where we factored out the shift in $\ket{n - 1}$ in $e^{-ik}$. Performing the sums over $n$ and $k'$, and further simplifying,

\begin{flalign}
\begin{aligned}
    & H_{BdG} = -\sum_k (w \sigma_z (e^{-ik} + e^{ik})   \\ & + i \Delta \sigma_y (e^{-ik} - e^{ik})) \ket{k}\bra{k} + \mu \sigma_z \ket{k}\bra{k}
    \\ & = - \sum_k [(\mu + 2w \cos(k))\sigma_z  - 2\Delta \sin(k) \sigma_y ]\ket{k}\bra{k}.
\end{aligned}
\end{flalign}

As such, we can write $H(k) = \bra{k} H_{BdG} \ket{k} = - (\mu + 2w \cos(k)) \sigma_z + 2\Delta \sin(k) \sigma_y$, where the full Hamiltonian is a sum of these $2 \times 2$ blocks over k. This can be diagonalized by setting $\operatorname{det}(H - E\operatorname{I}) = 0$ to obtain the band structure

\begin{equation}
    E(k) = \pm \sqrt{(2w\operatorname{cos}(k) + \mu)^2 + 4 \Delta^2 \operatorname{sin}^2 k}.
\end{equation}

As expected, there are two branches to the energy spectrum symmetric about $0$ and no zero-energy modes (as we closed the boundaries). The two branches touch at $k = 0$ when $\mu = -2w$, which, as we saw, is the transition between the two phases of the model.

At this point, we can jump back to a discussion on the open wire with modes on both edges. The modes are stable because, at $\mu = 0$, the Majoranas modes are localized at the end of the wire and separated by a gapped bulk. As a result, moving each one of these levels from zero energy would violate particle-hole symmetry. The only way to break the Majorana zero modes would be to couple them to each other, which is nominally forbidden by their spatial separation and the bulk gap. The splitting at $\mu \approx 2 w$ only occurs as the bulk energy gap is closed, merging the two edge modes.

Similarly, it is now also clear why these modes have zero energy. Since we have doubled the degrees of freedom by introducing holes, each pair of $\pm E$ actually corresponds to a single superposition of electrons and holes: a quasiparticle $b = u a + v a^\dagger$. An excitation far above the energy gap behaves similar to an electron, that is, $u \approx 0, v \approx 1$, and an excitation far below the energy gap behaves like a hole with $u \approx 1, v \approx 0$. Recall that the Majorana mode we are looking for can be written as an equal superposition of $a$ and $a^\dagger$ with $u^2 = v^2$: as expected, it should reside precisely in the middle of the gap, which is symmetric around $E = 0$.

We can now look more closely at what happens when the bulk closes, which marks the transition between the topological and trivial phases. We can approximate the Hamiltonian around the closing (where $k = 0$) to first order as

\begin{equation}
    H(k) \approx m\sigma_z + 2\Delta k \sigma_y,
\end{equation}

where $m = - \mu - 2w$ is the size of the energy gap, with an energy spectrum $E(k) = \pm \sqrt{(m^2 + 4 \Delta^2 k^2)}$. The sign of this mass term corresponds to the two phases of the bulk energy spectrum: When $m < 0$, $\mu > -2w$, the system is in the \textit{topological phase} where MZMs exist in an open chain, and when $m > 0$, $\mu < -2w$, it is in the \textit{trivial phase}. Note that when $m = 0$, the Hamiltonian reduces to $- 2\Delta \sigma_y i \partial_x$ with eigenenergies $E = \pm 2\Delta k$, corresponding to an equal superposition of electrons and holes --- Majorana zero modes. 

Let us look at this more closely by considering the case where the mass parameter $m$ varies in space and changes sign at $m(x = 0)$, such that there is a domain wall located at the origin and $m(x) \to \pm m$ as $x \to \pm \infty$. As we discussed, there are Majoranas located at the point where $m(x = 0) = 0$ with zero energy, so we can set $H\ket{\psi}$ to $0$:

\begin{equation}
    (m(x) \sigma_z - 2\Delta \sigma_y i \partial_x) \psi(x) = 0 ,
\end{equation}

rewriting $k = - i \partial_x$. Collecting the terms and using $ \sigma_z \sigma_x^{-1} = \sigma_y$, we obtain $\partial_x \psi(x) = m(x) \psi(x) \sigma_y/(2i\Delta)$, which can be solved by exponentiating the integral $\int \sigma_x m(x)/(2\Delta) dx$. This yields two solutions given by the eigenstates of $\sigma_x$:

\begin{equation}
    \psi(x) = e^{\pm \int^{x}_{0} \frac{m(x')}{2\Delta} dx'}\begin{pmatrix} 1 \\ \pm 1 \end{pmatrix}.
\end{equation}

Only one of these solutions is normalizable depending on whether $m(x)$ goes from positive to negative or vice versa. The solution represents a bound state localized at the domain wall where the wavefunction decays exponentially on both sides. This is the same exponential localization on the ends of the Kitaev wire that we see in Fig.~\ref{fig:spectrum} (b).

The chain is in the topological phase on the side where $m(x) > 0$, and in the trivial phase on the side where $m(x) < 0$: a zero-energy mode occurs at the interface of the two regions. This is what we might logically expect given that the trivial phase has Majorana modes paired by fermionic sites while the topological phase has the pairing offset by one. There will be an extra unpaired Majorana mode where the two phases meet.

\subsection{General conditions for MZMs}
\label{subsec:conditions}

One of the most interesting features of the Kitaev chain is its bulk topological invariant, which enables us to produce a general condition for the presence of unpaired Majoranas in a one-dimensional, translationally invariant Hamiltonian. Kitaev termed this invariant the Majorana number $\mathcal{M}(H) = \pm 1$, where the existence of MZMs is indicated by $\mathcal{M} = 1$. While its precise derivation is beyond the scope of this paper, we will now provide a heuristic explanation for this quantity which makes use of our finding on the gap closings.

As we discussed in Sec.~\ref{subsec:stability}, particle-hole symmetry ensures that for every eigenenergy at $E(k)$, there is also an eigenenergy at $-E(-k)$. Note that $k$, as defined, belongs in the range $(-\pi, \pi]$, so there are actually two exceptions, $k = 0$ and $k = \pi$, that map onto themselves. These happen to be the two points where the energy gap closes in momentum space. The gap closings correspond to fermionic parity switches: as the bulk spectra, corresponding to the Bogoliubov quasiparticles discussed in Sec.~\ref{subsec:stability} change signs, it becomes energetically favorable to add or remove a single quasiparticle. This changes the fermionic parity in the ground state from odd to even, or vice versa. We would like to have some way of counting these crossings, as it would provide a nice indication of whether the Hamiltonian can host MZMs.

It turns out that there is a clean mathematical formalism that allows us to create this exact invariant. We first perform a basis transformation on the Hamiltonian such that it becomes an antisymmetric matrix, where eigenvalues come in pairs (just like our eigenenergies). Since our Hamiltonian contains $\sigma_y$ and $\sigma_z$ terms, we can define the unitary matrix

\begin{equation}
    U = e^{- i \pi \sigma_y / 4} = \begin{pmatrix} 1 & 1 \\ i & -i \end{pmatrix},
\end{equation}

to perform a $\pi/2$ rotation around the y-axis onto the x-y plane. We can write the Hamiltonian at $0$ and $\pi$ as 

\begin{equation}
    \begin{aligned}
    H(0) = \frac{1}{2} U \begin{pmatrix} -2w-\mu & 0 \\ 0 & 2w + \mu \end{pmatrix} U^{\dagger} \\ = -i \begin{pmatrix} 0 & -2w-\mu \\ 2w+\mu & 0 \end{pmatrix},
    \end{aligned}
\end{equation}

\begin{equation}
    \begin{aligned}
    H(\pi) = \frac{1}{2} U \begin{pmatrix} 2w-\mu & 0 \\ 0 & - 2w + \mu \end{pmatrix} U^{\dagger} \\ = -i \begin{pmatrix} 0 & 2w-\mu \\ -2w+\mu & 0 \end{pmatrix}.
    \end{aligned}
\end{equation}

Note that for a general antisymmetric matrix, its determinant is always equal to the product of eigenvalues $\prod_n (-E_n^2)$. We would like to find another quantity where the square root of this determinant, $\pm \prod_n i E_n$, has a uniquely defined sign such that we can detect when a single $E_n$ changes with the fermionic parity. This is conveniently supplied by the Pfaffian of the matrix, which is defined as 

\begin{equation}
    \operatorname{Pf} = \frac{1}{2^N N!} \sum_{\tau \in S_{2N}} \operatorname{sgn}(\tau) A_{\tau(1), \tau(2)} \ldots A_{\tau_{(2N - 1)}, \tau_{(2N)}}.
\end{equation}

We can compute this quantity for our Hamiltonian at the two points, which yield $\operatorname{Pf}[iH(0)] = -2w - \mu$ and $\operatorname{Pf}[iH(\pi)] = 2w - \mu$. Each Pfaffian accounts for one of the gap closings that can occur in the bulk, so we can combine them to make a single invariant 

\begin{equation}
    \mathcal{M} = \operatorname{sgn(\operatorname{Pf}[iH(0)]\operatorname{Pf}[iH(\pi)])},
\end{equation}

which is the Majorana number which we were looking for. The condition that $\mathcal{M} = -1$ essentially states that for a Majorana mode to occur, the two states $H(0)$ and $H(\pi)$ must differ in their fermionic parity. Hence, we can write 

\begin{equation}
    \mathcal{M}(H_0) = (-1)^{\nu(\pi) - \nu(0)},
\end{equation}

where $\nu(k)$ counts the number of negative eigenvalues of the antisymmetrized $H(k)$. For $\mathcal{M} = -1$, we must have $ \nu(\pi) - \nu(0) = 1 \mod 2$, which hints at the challenge of implementing such a system: electron spectra are usually degenerate with respect to spin, so $\nu(\pi)$ and $\nu(0)$ are both even. The problem of spin degeneracy was actually hinted at when we discussed MZMs in the context of Bogoliubov quasiparticles $b = u a + v a^\dagger$ with $u^2 = v^2$. Actual electrons are described by spinful creation and annhilation operators $c_{\uparrow}$ and $c^{\dagger}_{\downarrow}$, which would not satisfy the defining relation of Majorana modes, $\gamma = \gamma^\dagger$. Potential solutions implementing such a spinless Hamiltonian, such as achieving p-wave pairing in much more readily available s-wave superconductors using an external magnetic field, were first proposed by several papers around 2010 \cite{oreg_helical_2010, lutchyn_majorana_2010}, and are still the subject of extensive experimental studies today.

\section{Non-abelian statistics}
\label{sec:non-abelian}

\begin{figure}
\includegraphics[width=0.5\textwidth]{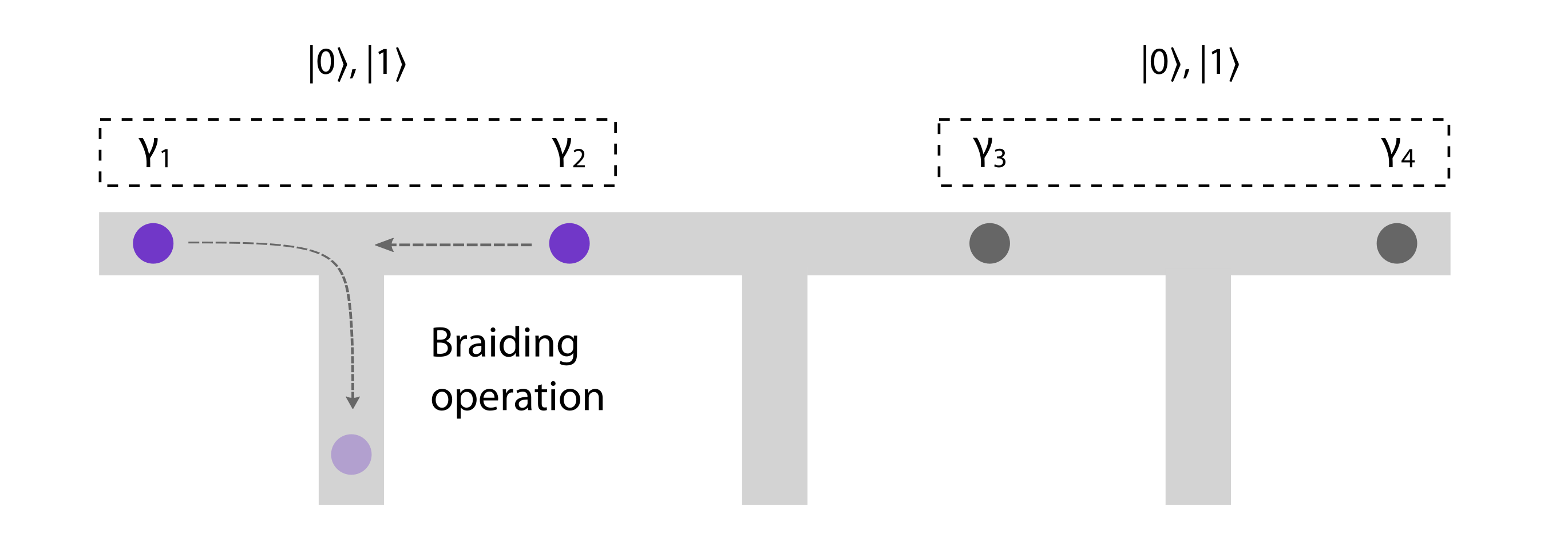}
\caption{\label{fig:braid} Braid operations on a network of network of nanowires hosting MZMs. Since the MZMs will fuse if they collide, in order to exchange $\gamma_1$ and $\gamma_2$, we need to first move $\gamma_1$ down, move $\gamma_2$ to the original position of the first mode, and then shuffle $\gamma_1$ to the position of the second mode. Here, each pair of MZMs encodes one fermionic state, which can be either $\ket{0}$ or $\ket{1}$.}
\end{figure}

Having explored how MZMs arise and form computational basis states, we can now discuss the exchange statistics of these modes. While swapping identical fermions or bosons yield $\ket{\psi_1 \psi_2} = - \ket{\psi_2 \psi_1}$ and $\ket{\psi_1 \psi_2} = \ket{\psi_2 \psi_1}$ respectively, the exchange of two MZMs twice does not leave the system invariant. It does not even follow the statistics of abelian anyons, where the exchange is described by an arbitrary phase factor $e^{i\theta} \ket{\psi_2 \psi_1}$, as the MZMs undergo a rotation in the degenerate ground state subspace that generally do not commute due to their high-dimensionality. This enables quantum information to be encoded by the braiding of Majorana modes, whose topological properties are stable under small perturbations \cite{freedman_topological_2002, alicea_non-abelian_2011}.

We need to go to a network of nanowires hosting 2N modes $\gamma_1 \ldots \gamma_{2N}$ \cite{baranov_lecture_nodate} to see the exchange statistics as braiding cannot occur in a single 1D wire. The modes are identical except for their positions in the network, forming a degenerate ground state manifold at zero energy. To number these states, we can group each neighboring pair of MZMs together into fermionic modes described by the parity operators $P_n = i \gamma_{2n-1}\gamma_{2n}$. Note that all the $P_n$ mutually commute as they correspond to individual fermions that do not share $\gamma_i$. We can use this set of parity operators to write our Hilbert space as

\begin{equation}
    \ket{\Psi} = \sum_{s_n \in {0,1}} \alpha_{s_1 s_2 \ldots s_N} \ket{s_1, s_2 \ldots, s_N},
\end{equation}

yielding a Hilbert space spanned by $\ket{s_1} \otimes \ket{s_2} \ldots \otimes \ket{s_N}$ with $2^N$ possible states. 

Now, suppose we swap the positions of two Majoranas $\gamma_i$ and $\gamma_j$ in the network. The evolution of the quantum state is given by $\ket{\Psi} \to U\ket{\Psi}$, where $U$ is some $2N \times 2N$ unitary matrix. There are two important conditions: first, the unitary evolution should only depend on the two Majoranas being exchanged, $\gamma_i$ and $\gamma_j$, and second, it must preserve the fermionic party of the system. As we argued for the quadratic Hamiltonian in Sec.~\ref{sec:kitaev_chain}, this means that the exchange must depend on their product $\gamma_i \gamma_j$. Thus, we can map the Hermitian product into a unitary operator by exponentiating it as $U = e^{\beta \gamma_i \gamma_j}$, up to some constant $\beta$. One of the Majorana operators $\gamma_i$ is transformed as

\begin{flalign}
    \begin{aligned}
    & U \gamma_i U^\dagger = \gamma_i + \beta [\gamma_i\gamma_j, \gamma_i] + \frac{\beta^2}{2!}[\gamma_i\gamma_j, [\gamma_i\gamma_j, \gamma_i]] \ldots \\ & = \gamma_i + \beta(\gamma_i \gamma_j \gamma_i - \gamma_j) + \frac{\beta^2}{2!}\gamma_i(\gamma_j \gamma_i \gamma_j \gamma_i + \gamma_j \gamma_i) \ldots \\ & = \gamma_i - 2\beta \gamma_j - \frac{(2\beta)^2}{2!}\gamma_i \ldots = \operatorname{cos}(2 \beta) \gamma_i - \operatorname{sin}(2\beta)\gamma_j,
    \end{aligned}
\end{flalign}


where we expanded the product using Hadamard's lemma. Similarly, $\gamma_j$ transforms into $U^\dagger \gamma_j U = \operatorname{cos}(2 \beta) \gamma_j + \operatorname{sin}(2\beta)\gamma_i$. Since we want $U$ to exchange the positions of $\gamma_i$ and $\gamma_j$, this constraints $\beta = \pm \pi/4$, which gives us the exchange operator for $\gamma_i$ and $\gamma_j$, 

\begin{equation}
    U = e^{ \pm \frac{\pi}{4} \gamma_i \gamma_j} =  \frac{1}{\sqrt{2}} (1 \pm \gamma_i \gamma_j).
\end{equation}

This operator acts nontrivially on the wavefunction beyond adding an overall phase and can, for example, place the system into a superposition of fermionic basis states $\ket{s_1} \ldots \otimes \ket{s_N}$. Furthermore, these exchange operators do not commute in general, e.g. $U_{ij} U_{jk} \neq U_{jk} U_{ij}$, making them non-abelian. 

\section{Discussion}

In our exploration of MZMs and the Kitaev chain, we encountered a wide range of important concepts in condensed matter physics, such as the Bogoliubov transformation, topological phases of matter, the bulk-edge correspondence, and non-abelian exchange statistics, all of which lead themselves to extremely rich and interesting discussions. Beyond these theoretical implications, Majorana modes also have the potential to revolutionize quantum information processing by providing a platform for decoherence-free information storage and manipulation. While a confirmed experimental realization of Majorana zero modes still eludes us \cite{castelvecchi_evidence_2021}, the physical implementation of such a system may provide an elegant solution to the problem of quantum computing, much like how the field-effect transistor --- a clever, highly scalable solution to building electrical switches --- ushered in a revolution in classical computing decades ago. 

\begin{acknowledgments}

The author would like to thank Seth Musser for extensive and helpful comments during the revision process, and Professor Maxim Metlitski for assistance with defining the scope of this paper. The organizational structure of this article is very much in debt to the Topology in Condensed Matter course developed by Akhmerov et al. \cite{top_cond}.

\end{acknowledgments}
\bibliography{references}

\end{document}